# scientific **data**

OPEN

DATA DESCRIPTOR

# A database of physical therapy exercises with variability of execution collected by wearable sensors

Sara García-de-Villa  ✉, Ana Jiménez-Martín  & Juan Jesús García-Domínguez

This document introduces the PHYTMO database, which contains data from physical therapies recorded with inertial sensors, including information from an optical reference system. PHYTMO includes the recording of 30 volunteers, aged between 20 and 70 years old. A total amount of 6 exercises and 3 gait variations were recorded. The volunteers performed two series with a minimum of 8 repetitions in each one. PHYTMO includes magneto-inertial data, together with a highly accurate location and orientation in the 3D space provided by the optical system. The files were stored in CSV format to ensure its usability. The aim of this dataset is the availability of data for two main purposes: the analysis of techniques for the identification and evaluation of exercises using inertial sensors and the validation of inertial sensor-based algorithms for human motion monitoring. Furthermore, the database stores enough data to apply Machine Learning-based algorithms. The participants' age range is large enough to establish age-based metrics for the exercises evaluation or the study of differences in motions between different groups.

## Background & Summary

Home health care is becoming more and more necessary, and not only for patients that have to follow a particular therapy after a surgical procedure, but also for people that require long-term therapies for a comprehensive strengthening, as happens with older adults. This is one of the pillars of active aging, where a set of exercises is prescribed to slow down the effects of frailty[1] and improve patients' functional ability[2]. Patients achieve the benefits of performing these therapies only when they remain consistent[3]. There is evidence that regular follow-up by health-care professionals keep patients active over the long term[4]. However, it is difficult to have a carer at the patient's disposal. This explains why virtual coaches are increasingly needed as an alternative to health staff, that could contribute to patients' adherence to physical exercise-based treatments[5]. In the absence of a personal supervisor, the system that works as a virtual coach has to perform three main tasks. Firstly, it has to get information about human motion, based on sensory systems, such as optical or inertial ones, among others. Secondly, it has to monitor physical activity and evaluate the performance of the exercises[6,7], what requires the development of particular algorithms adapted to physical therapy exercises[8–10], some of them based on Machine Learning (ML) methods[10,11], which relate parameters of human motions with the performance. And finally, the virtual coach has to provide feedback to the patient[5].

For the evaluation and optimization of algorithms, it is necessary particular datasets adapted to the exercises to be analysed. In this way, recent research efforts have focused on the creation of datasets by using different sensory systems. Video-based and portable technologies are the main alternatives in the monitoring of human activities[12], so databases obtained with both systems are of great interest. However, video-based technologies entail occlusions and patients' privacy concerns[13,14]. Conversely, wearable systems, such as inertial measurement units (IMUs), are becoming increasingly popular because of its practicality and its everywhere usable potential[15]. IMU-based databases of human motion monitoring commonly focus on the study and assessment of gait[16] or the study of activities of daily living[17], and they usually also include depth and RGB sensors.

This data descriptor aims to contribute to the research of the monitoring and evaluation of the performance of prescribed physical therapy exercises through inertial wearable sensors. This database, called PHYTMO (from

University of Alcala, Department of Electronics, Alcalá de Henares, 28801, Spain. ✉e-mail: sara.garciavilla@uah.es

| Range | Id | Age (years) | Height (cm) | Weight (kg) | Sex (M/F) | Motor conditions |
|---|---|---|---|---|---|---|
| A | A01 | 22 | 165 | 58 | F | Not reported |
| | A02 | 26 | 167 | 64 | F | Not reported |
| | A03 | 25 | 166 | 56 | F | Not reported |
| | A04 | 23 | 180 | 72 | M | Not reported |
| | A05 | 22 | 185 | 71 | M | Not reported |
| | A06 | 26 | 171 | 72 | M | Not reported |
| | A07 | 25 | 175 | 78 | M | Not reported |
| | A08 | 22 | 175 | 72 | M | Not reported |
| | A09 | 23 | 175 | 60 | M | Not reported |
| | A10 | 25 | 167 | 65 | M | Not reported |
| B | B01 | 30 | 179 | 76 | M | Not reported |
| | B02 | 34 | 185 | 84 | M | Not reported |
| | B03 | 39 | 161 | 51 | F | Not reported |
| | B04 | 31 | 164 | 58 | F | Not reported |
| | B05 | 38 | 176 | 59 | M | Not reported |
| C | C01 | 49 | 166 | 66 | F | Not reported |
| | C02 | 46 | 178 | 69 | M | Not reported |
| | C03 | 42 | 172 | 93 | M | Not reported |
| | C04 | 44 | 170 | 75 | F | Not reported |
| | C05 | 48 | 167 | 62 | F | Not reported |
| D | D01 | 50 | 168 | 72 | M | Not reported |
| | D02 | 56 | 172 | 85 | M | Not reported |
| | D03 | 51 | 154 | 67 | F | Not reported |
| | D04 | 54 | 160 | 62 | F | Not reported |
| | D05 | 55 | 165 | 75 | M | Pain in the right shoulder |
| E | E01 | 63 | 186 | 87 | M | Not reported |
| | E02 | 60 | 157 | 56 | F | Not reported |
| | E03 | 60 | 158 | 76 | F | Pain in the right shoulder and right knee |
| | E04 | 68 | 161 | 63 | M | Pain in the right shoulder |
| | E05 | 64 | 168 | 70 | F | Not reported |

**Table 1.** Anthropometric data, age and sex of volunteers. M and F stand for masculine and feminine, respectively.

PHYsical Therapy MOnitoring) is created for its use in the development of novel algorithms for the evaluation of human motions, including the identification and assessment of a known set of prescribed exercises. The database includes enough data for developing ML-based algorithms, as the authors proved in their proposal for the exercises recognition and evaluation that uses part of the inertial data of PHYTMO[18]. For the development of robust and generalizable algorithms, it is required a large amount of annotated data and the subjects variability is also important. PHYTMO includes data of the performance of 6 exercises and 3 gait variations commonly prescribed in physical therapies. Data are recorded with four IMUs placed on arms or legs, according to the performed exercise. Data also include the position and orientation of IMUs in the 3D-space during the performance of the 6 exercises measured with an optical system. Data of each exercise are divided into two kind of series, which consist in correctly and wrongly performed exercises. A total amount of 30 volunteers with variability in age and morphology performed these exercises, what makes it possible to study the differences between kinematic parameters that can occur in the execution of exercises at different ages. The anonymous subjects can be easily associated with their anthropometric information for this purpose. Furthermore, the data are labeled for the identification of each exercise separately, annotating its correct or incorrect performance.

Furthermore, this database can be very useful to train algorithms developed for human kinetic analysis. Using these data, IMU-based algorithms for kinetic parameter estimation can be developed or analyzed, as different proposals found in the literature[19–22]. This application is especially remarkable because PHYTMO include data from IMUs together with reference data from an accurate optical system, related to the 6 exercises based on repetitions of motions. Therefore, this database can be used to check different proposals using the same data, facilitating a fair comparison between algorithms.

## Methods

**Participants and ethical requirements.**    Thirty volunteers enrolled in the study:13 women and 17 men. Table 1 shows their anthropometric information together with their age, sex (masculine, M, or feminine, F) and their identifier (Id) in the database. Volunteers are separated by their age range in order to ease the analysis of different aged population. These ranges are clustered by decade, so range A includes volunteers aged between 20 and

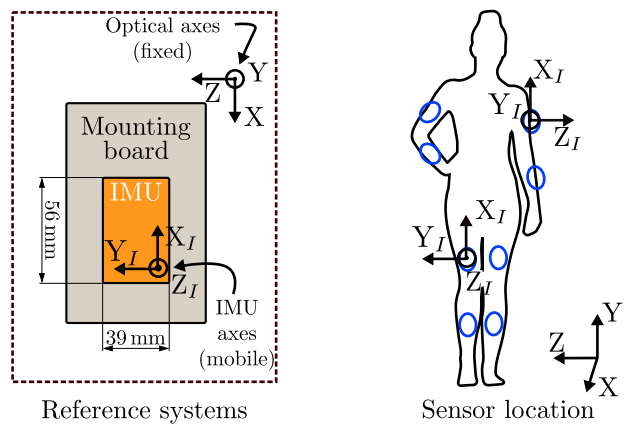

**Reference systems**        **Sensor location**

**Fig. 1** Axes of the optical and inertial systems and location of the IMUs on the lower-and upper-limbs. The picture on the left is a schema of the zenithal view of the IMU structure for the definition of its original orientation. The axes of the IMU and the optical systems are also depicted in the picture on the right, with the $X_I$-axes pointing to the ceiling in this position of the standing person. This picture also shows the axes of the optical system, where the XZ plane is parallel to the floor plane and the Y-axis points to the ceiling. The approximate location of IMUs (*Lshin*, *Lthigh*, *Rshin*, *Rthigh*, *Larm*, *Lforearm*, *Rarm* and *Rforearm*) are highlighted in blue in the picture. The silhouette graphic is from Vecteezy[51].

29 years old, B between 30 and 39, C between 40 and 49, D between 50 and 59 and E between 60 and 69. Table 1 also includes the volunteers' motor conditions that may have influenced in their movements. All volunteers were healthy and robust and only three of them reported pain during the motion, although they were able to perform all the evaluated exercises.

The study was carried out in the framework of FrailCheck project (SBPLY/17/180501/000392), following all the COVID-19 guidelines and recommendations. Volunteers wore masks during the exercises performance, which should be taken into consideration in the motion analysis because of possible early fatigue even in healthy volunteers. Guadalajara University Hospital approved the study protocol (Institutional Review Board No.2018.22.PR, protocol version V.1. dated 21/12/2020), and all participants signed a written informed consent.

**Acquisition setup.** The PHYTMO data set includes data recorded with four IMUs and an accurate optical system. These data were recorded in the Motion Capture Laboratory of the University of Alcala using the *NGIMU*[23] IMUs (*X-io Technology*, Bristol, UK) and the *OptiTrack*[24] system (NaturalPoint Inc). Wearable sensors include 3-axis gyroscope, accelerometer and magnetometer, with a range of 2000°/s, 16 g and 1,300 μT, respectively. These are common IMUs range values used for human motion monitoring, such as the popular *XSENS* sensors[16]. The sample rate was set to 100 Hz for the gyroscopes and accelerometers and to 20 Hz for the magnetometers in the reported data. The IMUs stored the recorded data on an SD card and as each volunteer finished performing the designed set of exercises, we downloaded the data to the computer for further data processing. They have a size of $56 \times 39 \times 18$ mm with a weight of 46 g, what makes them practical for wearing during the performance of exercises. Each IMU was mounted on an *ad-hoc* structure (mounting board) for its placing at the limbs.

Regarding the optical system, it is based on infrared light-emitting cameras situated in the capture room that identify the position of reflective markers placed on the subject anatomic landmarks and on the IMU structure. The optical system recorded the motions of the volunteers and the devices they wore along the data collection. We used the *OptiTrack* system, which consisted of eight depth *Prime 13* cameras with a resolution of 1.3 MP and a frame rate of 240 fps. We used this system with the *Motive 2.2.0* software to calibrate it before each use, to set the cameras rate to 100 Hz and to define the skeletons and objects corresponding to the IMU mounting boards to be recorded.

**Acquisition protocol.** The volunteer recordings were made in a continuous session on the day each volunteer was available and the individual sessions lasted an average of two hours. At the beginning of each day recordings, we set the coordinates origin of the optical system on the floor and always in the same point, so optical measurements are always referred to the same origin. We established the initial orientation of the IMU mounting board in the optical system placing this board on the floor, so in this orientation the rotation angles are equal to zero. The IMU $Y_I$-axis was parallel to the mounting board Z-axis and the IMU $Z_I$-axis was parallel to the board Y-axis, so the IMU $X_I$-axis was anti-parallel to the board X-axis, as depicted in Fig. 1-left.

The four IMUs were placed at the upper-or lower-limbs, according to the performed exercises, with their $X_I$-axis pointing to the ceiling, as the reference systems shown in Fig. 1-right. On the volunteers' lower-limbs, the IMUs were placed on the anterior surface, so when volunteers were standing, the $Z_I$-axis was perpendicular to the coronal plane of their bodies and the $Y_I$-axis was perpendicular to their sagittal plane (Fig. 1-right depicts these positions). On their upper-limbs, IMUs were placed on the exterior lateral position. In this case, when the volunteers' hands pointed to the floor, the $Y_I$-axis was perpendicular to the volunteers' coronal plane and the

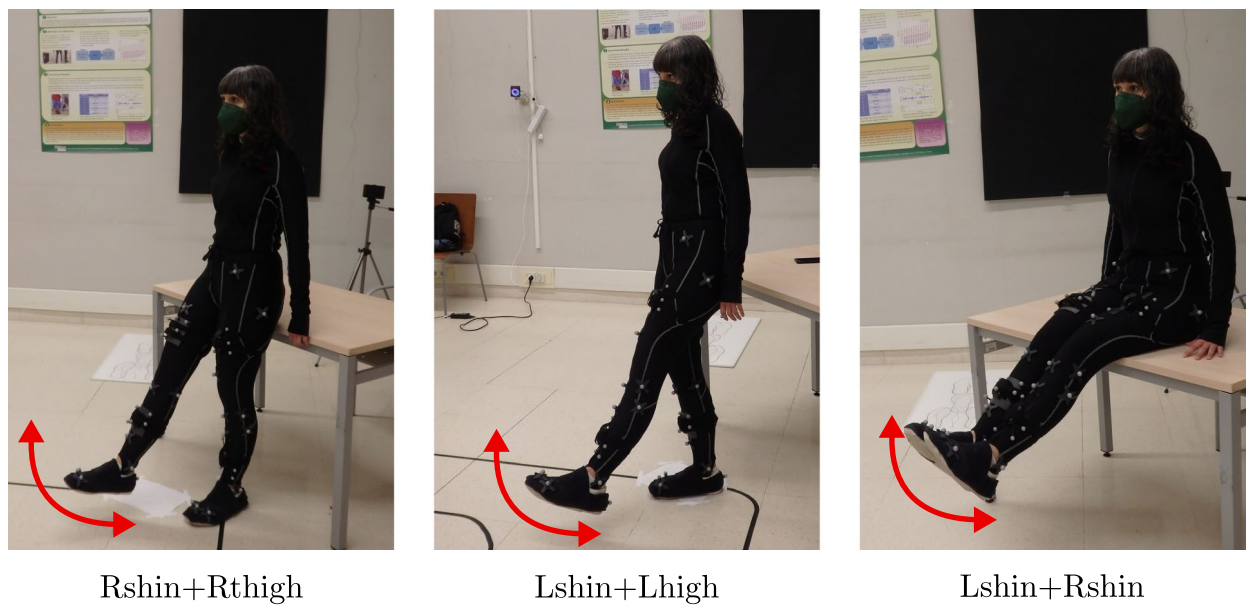

Rshin+Rthigh        Lshin+Lhigh        Lshin+Rshin

**Fig. 2** Data synchronization movements for the leg exercises. The labels under the pictures refer to the IMUs that are synchronized with each specific motion. The volunteer in the pictures has granted permission to publish.

$Z_l$-axis was perpendicular to their sagittal plane. We chose these locations on the body for the easiness of placing sensors. Each IMU has an identifier with format *Xsegment*, where *X* refers to its position, left ("L") or right ("R"), being *segment* "thigh" or "shin", referring to the segment of the lower-limb the IMU was placed on, or "arm" or "forearm" for the upper-limb.

We synchronized the four IMUs through the identification of significant events in specific motions at the beginning of each recording. For the synchronization of the leg exercises, volunteers performed the three motions depicted in Fig. 2, which consisted in: keeping the leg straight, two hip flex-extensions with the right leg to synchronize the two IMUs on this leg (Rshin and Rthigh), two hip flex-extensions with the left leg to synchronize the other two IMUs (Lshin and Lthigh), and two knee bending with both feet together in order to synchronize both legs by the detection of peaks in the signals of turn rate recorded by Lshin and Rshin. Data of arm exercises included two repetitions of straight arms elevation maintaining both hands together during the motion.

After the synchronization phase, volunteers performed the different exercises or gait variations, commonly prescribed in a specific way in physical therapies. These particular movements have been chosen as part of a physical exercise routine prescribed for elderly people to maintain their functional capacity[25]. These exercises have been tested in some studies[26,27], which demonstrate functional improvement in older adults after continued performance of them over a relatively short period of time (approximately 3 months). The exercises can be divided into two groups of three exercises: focused on the lower-limbs and on the upper-limbs. Table 2 lists all the exercises and gait variations carried out by the volunteers, and describes the correct way to perform them. Before the motion recording, an initial description of the exercises set was explained by one of the authors of the study, who also performed each motion as demonstration. Besides the correct performance of the exercises, volunteers made them wrongly, receiving no instructions in this case. However, most of volunteers performed the wrong repetitions of the exercises in similar way, so we also report the most common deviations during the wrong performances in Table 2. The quality of the performance was evaluated by an expert person, who indicated how the proper exercises had to be done and labeled the exercises as correctly or wrongly performed. We only consider one kind of wrong gait for the three variations (GAT, GIS and GHT) during which volunteers walk freely but pretending to be tired, dragging their feet on the floor. Figure 3 shows pictures of all these evaluated exercises properly performed.

The number of repetitions varies according to the age of volunteers, being the volunteers included in A and B ranges who performed the highest amount of repetitions. Volunteers made four times each exercise series. Two of these series consisted in the corresponding exercises properly performed and, in the other two series, the exercises were wrongly done.

**Calibration of legs.** In the human motion analysis field, it is common to use the information of the joints location or orientation[19–21,28]. Thus, PHYTMO also includes three motions for the calibration of joints. The volunteers performed the following three motions in order to have information to calibrate their lower-limbs:

- Hip circles: standing, keeping the hip still and maintaining the legs completely straight, the volunteers performed circles with one leg, with the center of rotation (COR) of hips stable (see Hip COR in Fig. 4). With this motion, the COR can be estimated using algorithms as different works propose[29–31].

| Exercise | Correct execution | Wrong execution |
|---|---|---|
| Knee flex-extension (KFE) | Sat on a stable surface, from the initial position of 90° of knee flexion, keeping the left leg still the right one moves in the sagittal plane extending the knee until its maximum. This variation is labeled as KFER. After all repetitions moving the right leg, this one remains still and the left one moves (KFEL). Number of repetitions: 10–20. | • Deviations from the sagittal plane<br>• Focusing the motion on the thigh<br>• Moving both legs instead of only one of them |
| Squats (SQT) | From standing position, volunteers make the exercise sitting on a chair, avoiding the lateral bending of knees or hip, and when touching the chair they stand up again. Number of repetitions: 8–15. | • Seating on the chair<br>• Without touching the chair<br>• Supporting their weight with their hands during the sitting or standing |
| Hip abduction (HAA) | Standing up, keeping the left leg still the right one moves upwards to the exterior side in the volunteer's frontal plane, remaining straight. This variation is labeled as HAAR. After all repetitions moving the right leg, the right leg remains still and the left moves (HAAL). Number of repetitions: 10–20. | • Deviations from the frontal plane<br>• Bending the knee during the execution<br>• Lack of motion control |
| Gait (GAT) | Volunteers walk freely in the room. Number of repetitions >20. | • Dragging the feet |
| Gait describing ∞ (GIS) | Volunteers walk around two objects on the floor, describing a trajectory similar to the infinity symbol (∞). Number of repetitions >20. | • Dragging the feet<br>• Without following the trajectory (GAT) |
| Gait with heel-tiptoe (GHT) | During walking, volunteers place first the heel on the floor and then they raise into their tiptoe. Keeping their weight into their tiptoe, they place the other heel on the floor and repeat the motion. Number of repetitions >20. | • Walking without the heel-tiptoe motion (GAT) |
| Elbow flex-extension (EFE) | Both arms move from the straight position to the maximum flexion of elbows in the sagittal plane, keeping the shoulders still. Number of repetitions: 10–20. | • Moving only one arm<br>• Focusing the force on the back<br>• Deviating the motion from the sagittal plane |
| Extension of arms over head (EAH) | With both hands together, arms, as straight as possible, make a arch until reaching the maximum elevation of hands. Number of repetitions: 10–20. | • Making the force with only one arm<br>• Not raising the arms over the head<br>• Separating both hands |
| Squeezing (SQZ) | Using a clothing and keeping arms straight forward, wrists move anti-symmetrically squeezing the clothing. Number of repetitions: 10–20. | • Moving only one wrist<br>• Turning wrists in other directions |

**Table 2.** Exercises included in the database together with the description of their correct and wrong performance.

- Hip frontal flex-extensions: standing, keeping the hip still and maintaining the legs completely straight again, the volunteers moved this leg in a forward-backwards motion (see Hip axis ⊥ sagittal in Fig. 4). With this motion, the axis perpendicular to the sagittal plane can be determined using different methods in the literature[32,33].
- Knee frontal flex-extensions: sat on a stable surface, the volunteers moved the shin of one leg from its knee in a forward-backwards motion with a low range of motion, around 30°, keeping the knee still (see Knee axis in Fig. 4). With this motion, the location of the axis can be determined using some proposed methods[31–33].

Similar motions were performed in order to calibrate the volunteers' upper-limbs, so the aforementioned methods, which are suggested for the inertial calibration of lower-limbs, can also be applied in this scenario. In this case, volunteers performed the following three motions:

- Shoulder circles: keeping the shoulder still and one arm completely straight, the volunteers performed circles with this arm (see Shoulder COR in Fig. 4).
- Shoulder frontal flex-extensions: keeping the shoulder still and maintaining one arm completely straight, the volunteers moved it in a forward-backwards motion (see Shoulder axis ⊥ sagittal in Fig. 4).
- Elbow frontal flex-extensions: keeping the elbow still and the hand in the supine position, the volunteers moved one forearm from its elbow in a forward-backwards motion with a low range of motion, around 60° (see Elbow axis in Fig. 4).

**Data processing.** As previously mentioned, we synchronized the four IMUs through the identification of significant events in the recorded signals during specific motions at the beginning of each recording. We used *MATLAB R2020b*[34] to manually select the time instants of negligible turn rate recorded with each sensor in order to set the initial time of each signal. We exported the data of each sensor separately in CSV format. More information about the name and organization of data is provided in the following section.

With respect to the optical system data, we used the *Motive 2.2.0* software[35] to fill gaps caused by occlusions during the data recording, so we provide the raw and interpolated optical data. Firstly, we corrected the mislabeled markers in their current location. Secondly, we interpolated using the *Model based* option, which uses the information of the visible markers of an object to infer the trajectory of the others. Finally, we used the *Cubic* interpolation in order to fill the information of markers in which the previous interpolation technique did not work because less than three markers of an object were seen. This technique uses a cubic spline to fill the missing data.

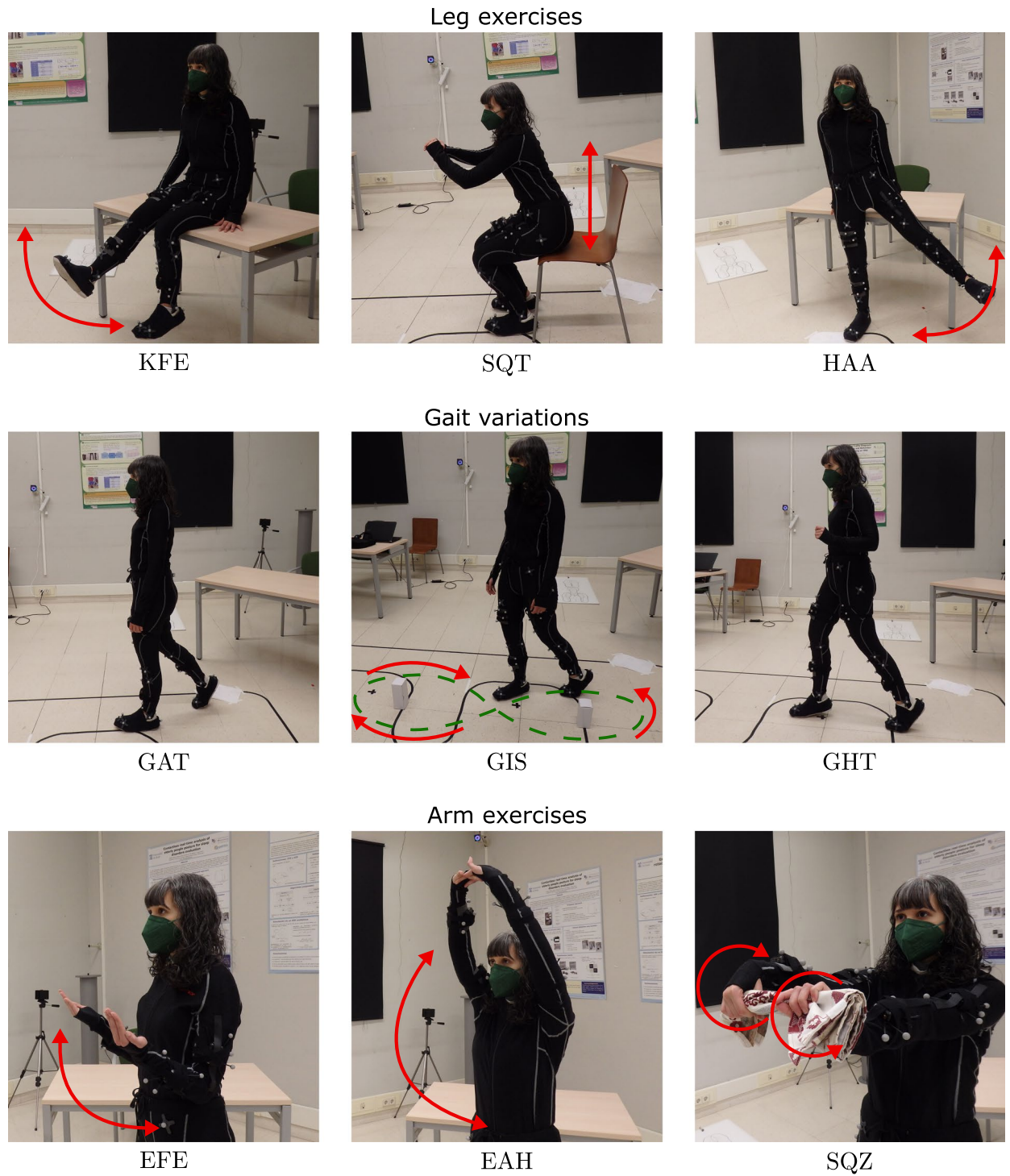

**Fig. 3** Exercises considered in this study performed by one of the volunteers. The first row includes knee flex-extension (KFE), squats (SQT) and hip abduction (HAA). The middle row contains natural gait (GAT), gait describing the infinity symbol (∞) in the trajectory (GIS) and heel-tiptoe gait (GHT). The last row presents elbow flex-extension (EFE), extension of arms over head (EAH) and squeezing (SQZ). PHYTMO includes the inertial data during all these exercises and the reference from a three-dimensional optical system of the leg and arm exercises (first and last rows). The volunteer in the pictures has granted permission to publish.

## Data Records

**Raw data.**  All raw data files exported from both inertial and optical systems were stored as CSV files and have been uploaded to Zenodo[36]. A total of 7,076 files are available with https://doi.org/10.5281/zenodo.6319979. Files are called with the nomenclature *GNNEEELP_S*, where *G* refers to the letter of the range of age, so it is "A", "B", "C", "D" or "E" (see Table 1); *NN* is number of identification of the volunteer, which ranges from "01" to "10"; *EEE* indicates the type of exercise (KFE, HAA, SQT, EAH, EFE or SQZ) or gait variation (GAT, GIS or GHT); *L* is the leg with the exercise is performed, so this letter is only included in the KFE and HAA exercises and it can be "L" or "R"; *P* is a label that indicates the evaluation of the exercise performance, which takes the "0" value when the file contains the correctly performed exercise and "1" when exercises are wrongly performed; and finally, *S* indicates the index of the series, being "1" for the first recorded series and "2" for the second one. In this way, if the third (03) recorded volunteer aged between 41 and 50 years old (C) performs the knee flex-extension (KFE) of the right leg (R) following the prescriptions of the exercise (0) in the first series (1), the corresponding file is called "C03KFER0_1.csv".

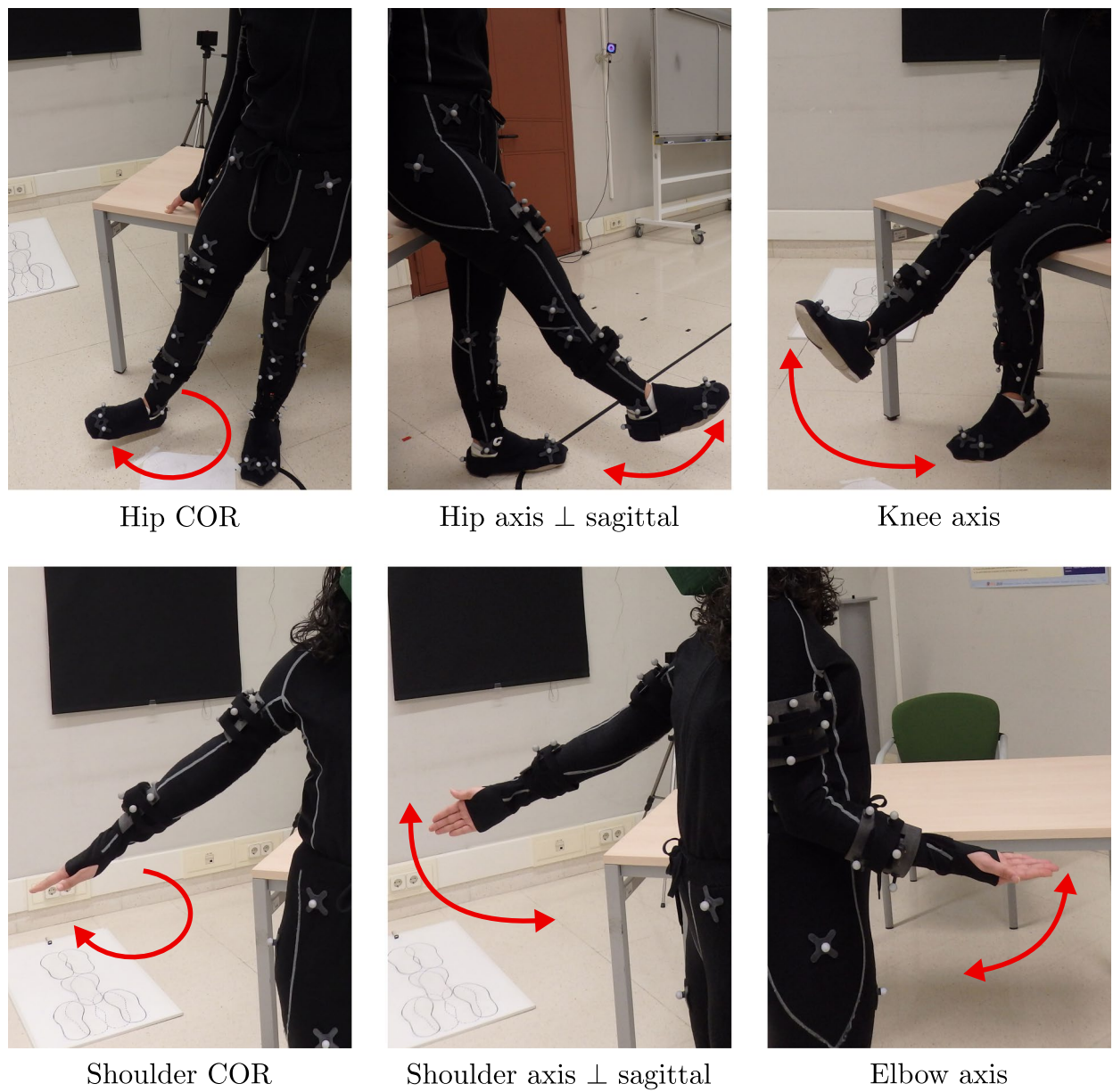

**Fig. 4** Motions performed for the calibration of the sensors with respect to the location and orientation of the CORs and axes of rotation of leg joints. Hip COR is the motion performed in order to determine the COR of hips, Hip axis ⊥ sagittal corresponds to the movements carried out to calibrate the hip axis perpendicular to the sagittal plane of the body and Knee axis is the motion that allows us to locate the knee axis and its orientation.

The main directory includes three folders called "inertial", "optical raw" and "optical interp", which contain the data recorded with the IMUs and with the optical system, respectively. The "raw" and the "interp" folders contain the raw and interpolated data, respectively. The data organization is different for the inertial and for the optical data, so they are separately explained bellow.

On the one hand, the "inertial" folder is divided into two directories which refer to the two possible group of limbs, that is "upper" or "lower". The corresponding internal structure is schematized in Fig. 5 and detailed in the following. Each limb directory contains five folders ("A", "B", "C", "D" and "E"), corresponding to each age group of volunteers. The age group folders contain one directory for each limb segment, which follows the names given

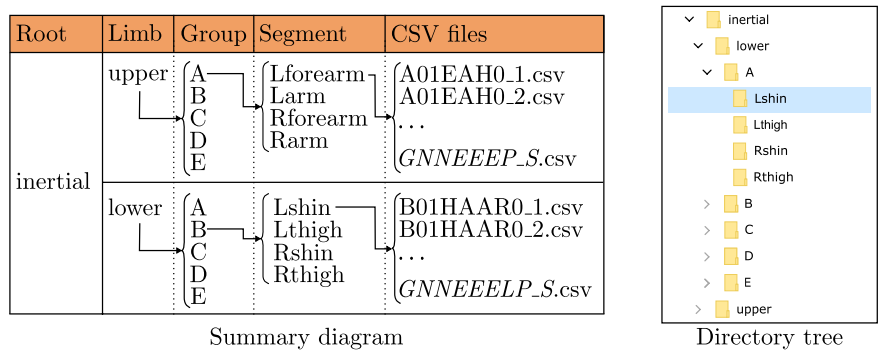

Summary diagram        Directory tree

**Fig. 5** Data organization at the inertial folders. The upper and lower folders are included in the "inertial" directory. The "upper" and "lower" terms refer to the upper-and lower-limbs. The five groups correspond to the five age groups in which volunteers are organized. The segment folders are called according to the body side in which the IMUs are placed (left-L or right-R), and the corresponding limb segment. Finally, the CSV files follow the *GNNEEELP_S* nomenclature (*G*: range of age of the volunteer, *NN*: number of its identification, *EEE*: type of exercise, *L*: leg that moves, only in KFE and HAA exercises, *P*: evaluation of the exercise performance and *S*: index of the series).

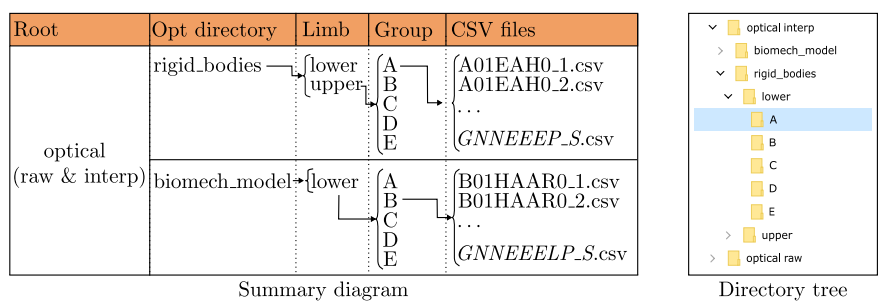

Summary diagram        Directory tree

**Fig. 6** Data organization at the optical folders. The upper and lower folders are included in the "optical raw" and "optical interpolated" directories. The "upper" and "lower" terms refer to the upper-and lower-limbs. The five groups correspond to the five age groups in which volunteers are organized. Finally, the CSV files follow the *GNNEEELP_S* nomenclature (*G*: range of age of the volunteer, *NN*: number of its identification, *EEE*: type of exercise, *L*: leg that moves, only in KFE and HAA exercises, *P*: evaluation of the exercise performance and *S*: index of the series). The names of the files with interpolated data follow the same nomenclature and have "_interp" at the end.

to the IMUs. Thus, there are four directories for the lower-limbs ("Lshin", "Lthig", "Rshin" and "Rthigh") and other four for the upper-limbs ("Lforearm", "Larm", "Rforearm" and "Rarm").

On the other hand, the other two folders, "optical raw" and "optical interp", contain two directories: "biomech_model" and "rigid_bodies", whose internal structure is schematized in Fig. 6. The data of the markers placed directly on the body of each subject are contained in the "biomech_model" folder and the data of the IMU mounting boards are located in the "rigid_bodies" folder. Thus, "biomech_model" has only one folder, called "lower" to indicate that it corresponds to the lower-limbs, and this folder contains the five directories of each age group. Conversely, the "rigid_bodies" folder is divided into the two directories "upper" and "lower", with the corresponding age group folders (see Fig. 6). Since the optical system includes the orientation and location of all the recorded objects in only one file, these data are not organized in segment folders but are included directly in the age group directories.

Inertial systems, whose final CSV files organization is indicated in Fig. 5, measure the turn rate, linear acceleration and magnetic field with the timestamp. This information is labeled in the inertial files as "Time (s)", "GyroscopeA (deg/s)", "AccelerometerA (g)" and "MagnetometerA (uT)", where "A" refers to the corresponding measurement axis (X, Y or Z). The information given by each sensor is detailed in Table 3.

Additionally, we provide the reference data of the optical system during the six repetitive exercises, i.e. the orientation and location in the 3D space of the IMU mounting boards. These files include three rows for the explanation of the recorded data. The first row indicates if the information corresponds to an IMU mounting board (Rigid Body) or to a marker (Rigid Body Marker). The second row contains the name of the corresponding Rigid Body. We defined four rigid bodies, named "shin" and "thigh" for the structures placed at the right side of the body and "shin2" and "thigh2" for the ones on the left side. The same mounting structures were used on legs and arms, placing always the "shin" named structures on the shin or forearm, and the "thigh" on the thigh or arm, according to the recorded exercise. In the Rigid Body Marker case, the second row indicates the number of marker as "Rigid Body: MarkerNum", e.g. "shin:Marker1" corresponds to the marker labeled as "1" by the optical system that is part of the Rigid Body "shin". Finally, the third row distinguishes between the Rotation and Position information. The Rotation of the Rigid Bodies with respect to their initial position (see Fig. 1) is provided in quaternions (note that markers do not have rotation information). The Position of the Rigid Bodies and the Rigid Body Markers, referred to the coordinates origin of the optical system, have magnitude of meters, as detailed in the "OBJ/position" and "OBJ/orientation" rows in Table 3.

| | Column label | Unit | Description |
|---|---|---|---|
| Inertial | Time | s | Time since the turn on of the device until the recording of each sample |
| | Gyroscope | °/s | Turn rate. Divided into three coordinates: X, Y and Z, which correspond to the vertical, lateral and anterior directions. |
| | Accelerometer | g | Linear acceleration with the influence of the gravity force. Divided into three coordinates: X, Y and Z, which correspond to the vertical, lateral and anterior directions. |
| | Magnetometer | $\mu T$ | 3D magnetic field. Divided into three coordinates: X, Y and Z, which correspond to the vertical, lateral and anterior directions. |
| Optical | OBJ/position | m | Position in/into the 3D space of the IMU object. Divided in three columns: X, Y and Z |
| | OBJ/orientation | N/A | Components of the orientation quaternion of the IMU object in the 3D space. Divided in four columns: X, Y, Z and W |
| | SKT/position | m | Position in the 3D space of the skeleton segment. Divided in three columns: X, Y and Z |
| | SKT/orientation | N/A | Components of the orientation quaternion of the joint in the 3D space. Divided in four columns: X, Y, Z and W |

**Table 3.** Label of columns in the CSV files together with their units and description.

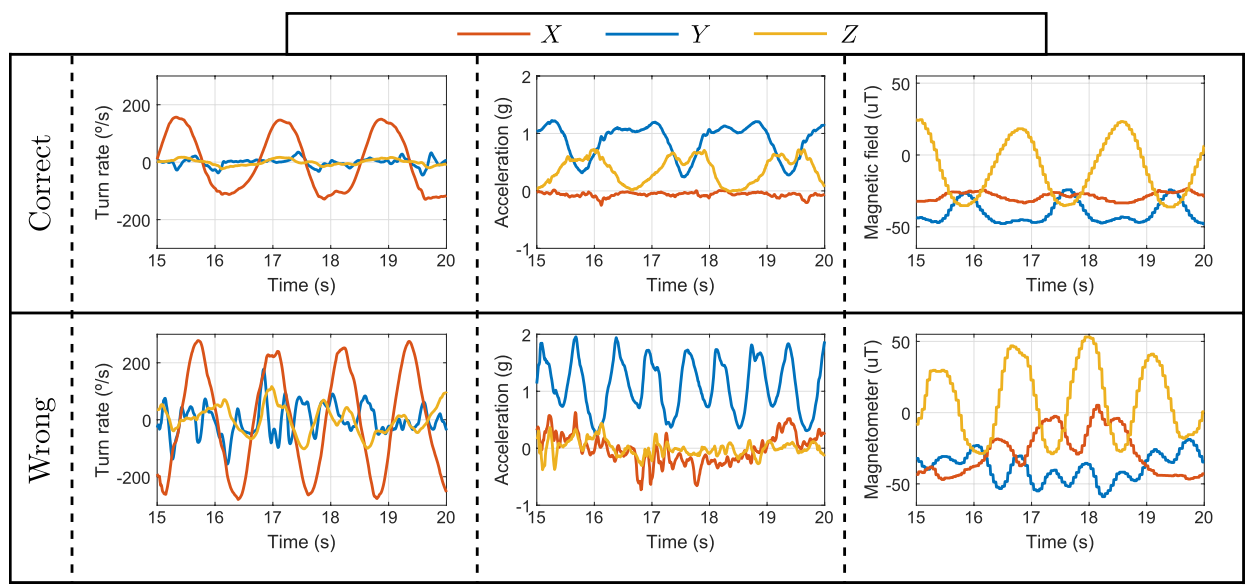

**Fig. 7** Signals from the gyroscope, accelerometer and magnetometer of *Lshin* during the KFEL exercise performed by volunteer A02. The corresponding files are "A02KFEL0_2.csv" and "A02KFEL1_1.csv", which are located at the "Lshin" folder.

PHYTMO also includes an extra file of the biomechanical model during the calibration of the lower-limbs of volunteers. In this way, we provide relevant anthropometric information that can be needed or used in the development of IMU-based algorithms for motion monitoring, as previous works propose[19–21,28]. The data of the lower-limb segments are indicated with the label "Bone" and when the data are referred to the position of markers, the corresponding label is "Bone Marker". We use the Rizzoli Lower Body Markerset[37], so the markers are called according to its protocol. The position of the Bones and the Bone Markers is measured in meters, as detailed in the "SKT/position" and "SKT/orientation" rows in Table 3. The names of skeletons are the group and identifier of the volunteer, as *GNN*, following the rules of the previously explained for the *GNNEEELP_S* nomenclature.

As an example, one representation of the inertial data during an exercise performed by one of the volunteers is depicted in Fig. 7. We show the signals obtained with the tri-axial gyroscope, accelerometer and magnetometer of *Lshin* during the KFEL exercise when it is correctly performed and when the volunteer performed it wrongly.

Figure 7 shows that the correctly and wrongly performed exercises are alike, following approximately similar patterns. However, the former is regular, the nine signals have similar amplitude in repetitions, whereas the latter shows changes between repetitions. Also, differences in the repetition duration and the magnitude of the signals can be observed. The features of these signals are studied in the following section in order to quantify these differences in performance.

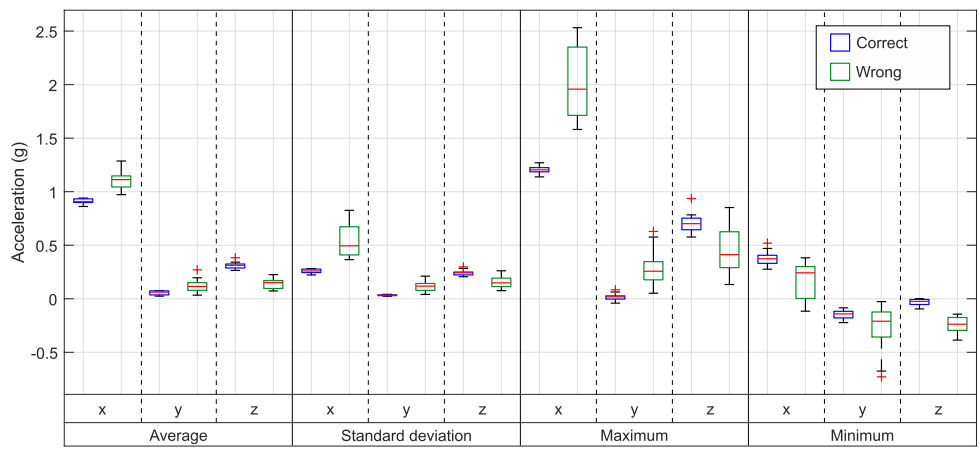

**Fig. 8** Boxplots of the average, standard deviation, maximum and minimum of the acceleration measured by *Lshin* during the KFEL exercise performed by volunteer A02.

**Processed data.** The files from the inertial system are provided in raw format, synchronized as previously explained. These signals contain relevant information for discerning the two kinds of performance of exercises: correct and wrong. One possible approach to extract information from the inertial system signals is to split them into repetitions of the exercises and analyze different features as the average, standard deviation, maximum and minimum of these segments.

One example of this analysis is represented in Fig. 8, which shows some features of the three signals obtained with the accelerometer in *Lshin* through *boxplots*. Boxplots are standardized for displaying features using five numbers: minimum, maximum (both excluding the outliers), median, first quartile and third quartile. The outliers are also commonly depicted beyond the maximum and minimum. Thus, the boxplots in Fig. 8 depict the quartiles and outliers of four features extracted from the acceleration signal of each repetition of the KFEL exercise correctly and wrongly performed by A02. The features analyzed with these boxplots are the average, standard deviation, maximum and minimum.

Figure 8 also shows the features of the correct exercises and the wrong ones. In this way, we can do not only a comparison between the two kinds of performance, but also study the dynamics during both performances. The correct exercises show a low dispersion of data, with smaller boxes than those which correspond to the wrong performance. That is consistent with the signals depicted in Fig. 7, which have similar amplitudes in the correct performance and high differences in its wrong performance. Furthermore, the median commonly differs between the two kinds of performance. Thus, the differences among these data make possible the classification between a correctly and wrongly performed exercise.

## Technical Validation

**Sensor placement.** Volunteers wore tight and sporty clothing for the experiments in order to prevent sensor movement and limitations in the exercises performance. As described in the Methods (see Methods/Participants and Fig. 3), the wearable sensors were placed on the legs or arms, according to the recording of an upper-or lower-limb exercise, and always with the $X_I$-axis pointed to the ceiling when volunteers were standing up. Sensors and optical markers were placed by the same two researchers to ensure consistency. It is also remarkable that previous to each recording, the optical system was calibrated, obtaining an error lower than 1 mm in all cases.

**Missing data.** Optical systems commonly loose data because of occlusions. We provide the raw and the filled data, in which we interpolate the missing data, in order to ease its use. The raw data include a 6.9% of occlusions and we reduce until a 1.3% of them when the data were interpolated. The interpolated files still have some missing data due to occlusions larger than 2 seconds in the raw data that were not accurately interpolated, as happened during the performance of the SQT exercise by some subjects who accidentally covered one or two IMUs during the exercise performance. We provide those files in which the data of some mounting boards are recorded as Rigid Bodies and other are missed. The recorded boards are given raw and interpolated along the complete recording.

When there were recording problems with the inertial system, we removed the erroneous data in PHYTMO (5%). As a consequence, some volunteers have less recorded exercises.

**Comparison with published data sets.** This database focuses on the study of the motion monitoring during the performance of physical therapies, which is a relevant topic of growing interest. This is a remarkable distinction, since the previous databases related with the human motion monitoring commonly study walk patterns and variations[16,38,39], activities of daily living[40–43] or are specific for different sports, such as football[44] or karate[45], but few data of prescribed exercises are publicly available.

Besides this lack of data, the existing physical exercises databases aimed for physical therapies are recorded with optical systems, such as three-dimensional systems and RGB together with depth sensors[46], or only with the RGB-depth systems[47]. This implies a limitation for the physical therapies that can only be performed in controlled environments where no occlusions occur. For the best of our knowledge, no databases of alternatives based on wearables are available. Regarding the number of participants, this database includes a total of 30 subjects whereas in previous studies of specific exercises less than 10 subjects were recorded. Thus, PHYTMO contains data for the analysis of the inertial-based monitoring of physical therapies with a high variability of participants. Related to this topic, data can be used in ML-based studies for classification purposes, such as the identification of a set of exercises[48]. As a matter of fact, this database has been used for the simultaneous recognition and evaluation of exercises in physical routines[18]. Another possible use is the study of motion kinematics, focused on the raw inertial data or aiming to evaluate joint angles, an important parameter during these prescribed exercises[46].

Since physical therapies also include different variations of gait, PHYTMO contributes to the study of walking patterns with inertial sensors. Natural gait and gait variations study is of great interest[16,38] and, as previously mentioned, databases focused on gait already exist, but they are commonly obtained only with optical sensors[38,39]. Conversely, we provide data obtained with magneto-inertial sensors, including 3 different variations of gait not available yet in previous datasets.

## Usage Notes

Different studies in the literature show that IMUs are valid for human motion monitoring, for the description of movements and for their evaluation. To ensure a huge usability of this database, the provided files are in CSV format, so they can be easily imported into Python or MATLAB.

This work contributes to the human motion analysis field by overcoming the lack of available data to compare algorithms. The validation of the proposals is so far performed on different data obtained for each of the studies. This difference makes it impossible to draw a fair comparison between them and, sometimes, not even the same metrics are available to compare them. This database helps to the development of new alternatives for motion analysis, by using the inertial data provided, and to their validation by using the reference data from the optical system. In this way, these data contribute to assess the strengths and limitations of various proposals. In addition, data are provided so that the research process does not require a previous step of collecting volunteers and recording movements in motion capture laboratories, and the data can be used directly.

Python tools are available, specifically for the analysis of the gait variations, which can be processed with GaityPy, which are Python functions to read accelerometry data and estimate gait features (https://pypi.org/project/gaitpy/). MATLAB also have tools for human motion analysis: the Kinematic and Inverse Dynamics toolbox (https://www.mathworks.com/matlabcentral/fileexchange/72863-forward-and-inverse-kinematics) can be used to study human kinematics and dynamics, and the Inertial Measurement Unit position calculator (https://www.mathworks.com/matlabcentral/fileexchange/25730-inertial-measurement-unit-position-calculator) can be applied for calculating the body's trajectory, velocity and attitude using data from an IMU as input.

Additionally, the presented IMU data can be used in OpenSim[49], a freely available tool for musculoskeletal modeling and dynamic simulation of motions. Specifically, OpenSim has a workflow called OpenSense (https://simtk-confluence.stanford.edu/display/OpenSim/OpenSense+-+Kinematics+with+IMU+Data) that details the corresponding data processing to simulate the kinematics of the body using the measurements of IMUs.

This database contributes in other human monitoring fields different from the motion analysis. The data include different exercises and subject variability, which are also labeled for different classification applications. Some of these classifications can be human activity recognition, motion evaluation, or optimization of localization and number of sensors.

It should be noted that the sensors have always been placed in the same body area, but not in a specific anatomical landmark. This is important to mention, since the data included cannot be used in studies of the kinematics of different populations, although they are useful for the development of movement analysis and monitoring algorithms, contributing in these motion monitoring areas.

Finally, we help to its usability in this classification problems with the publication of the *features_extraction* function, developed for MATLAB in any of its version. This function splits signals using a sliding window, returning its segments, and extract signal features, in the time and frequency domain, based on prior studies of the literature[6,10,11,48,50]. This function needs at least three signals from one triaxial sensor to extract their features, a window size in number of samples and a window shift that defines the distance between consecutive windows. The returning features are divided into the time and frequency domains. The time domain features are: mean of each signal, maximum of each signal, minimum of each signal, mean of the absolute value of each signal, standard deviation of each signal, variance of each signal, mean absolute deviation of each signal, root mean square of each signal, mean over the three axes of each sensor, average standard deviation over each three axes, skewness of each signal, average skewness over each three axes, kurtosis of each signal, average kurtosis over each three axes, 25, 50 and 75 quartiles of each signal, power of each signal, correlation between the three axes, correlation of each axis and the vector norm and entropy of each signal. The frequency domain features are always for each signal and they are the following: energy, absolute value of the maximum FFT coefficient, absolute value of the minimum FFT coefficient, maximum FFT coefficient, mean FFT coefficient, median FFT coefficient, discrete cosine transform and spectral entropy. Finally, a box plot of the selected features and sensor is depicted to ease the analysis of data in PHYTMO.

## Code availability

No custom code was used to generate or process the data.

## References

1. Rodríguez-Mañas, L., Rodríguez-Artalejo, F. & Sinclair, A. J. The third transition: the clinical evolution oriented to the contemporary older patient. *Journal of the American Medical Directors Association* **18**, 8–9 (2017).
2. Izquierdo, M., Duque, G. & Morley, J. E. Personal view physical activity guidelines for older people: knowledge gaps and future directions. *The Lancet Healthy Longevity* **2** (2021).
3. Jack, K., McLean, S. M., Moffett, J. K. & Gardiner, E. Barriers to treatment adherence in physiotherapy outpatient clinics: a systematic review. *Manual therapy* **15**, 220–228 (2010).
4. Bennett, J. A. & Winters-Stone, K. Motivating older adults to exercise: what works? (2011).
5. Kyriazakos, S. *et al.* A novel virtual coaching system based on personalized clinical pathways for rehabilitation of older adults-requirements and implementation plan of the vcare project. *Frontiers in Digital Health* **2** (2020).
6. Bavan, L., Surmacz, K., Beard, D., Mellon, S. & Rees, J. Adherence monitoring of rehabilitation exercise with inertial sensors: a clinical validation study. *Gait & Posture* **70**, 211–217 (2019).
7. Mancini, M. *et al.* Continuous monitoring of turning in Parkinson's disease: rehabilitation potential. *NeuroRehabilitation* **37**, 783–790 (2017).
8. Maciejasz, P., Eschweiler, J., Gerlach-Hahn, K., Jansen-Troy, A. & Leonhardt, S. A survey on robotic devices for upper limb rehabilitation. *Journal of NeuroEngineering and Rehabilitation* **11**, 1–29 (2014).
9. Gauthier, L. V. *et al.* Video game rehabilitation for outpatient stroke (vigorous): protocol for a multi-center comparative effectiveness trial of in-home gamified constraint-induced movement therapy for rehabilitation of chronic upper extremity hemiparesis. *BMC Neurology* **17**, 109 (2017).
10. Pereira, A., Folgado, D., Cotrim, R. & Sousa, I. Physiotherapy exercises evaluation using a combined approach based on sEMG and wearable inertial sensors. In *Proceedings of the 12th International Joint Conference on Biomedical Engineering Systems and Technologies - Volume 4: BIOSIGNALS*, 73–82 (SciTePress, 2019).
11. Zhao, L. & Chen, W. Detection and recognition of human body posture in motion based on sensor technology. *IEEJ Transactions on Electrical and Electronic Engineering* **15**, 766–770 (2020).
12. Cust, E. E., Sweeting, A. J., Ball, K. & Robertson, S. Machine and deep learning for sport-specific movement recognition: a systematic review of model development and performance. *Journal of sports sciences* **37**, 568–600 (2019).
13. Komukai, K. & Ohmura, R. Optimizing of the number and placements of wearable imus for automatic rehabilitation recording. In *Human Activity Sensing*, 3–15 (Springer, 2019).
14. Zihajehzadeh, S., Member, S., Park, E. J. & Member, S. A novel biomechanical model-aided IMU/UWB fusion for magnetometer-free lower body motion capture. *IEEE Transactions on systems, man and cybernetics: systems* 1–12 (2016).
15. Lopez-Nava, I. H. & Angelica, M. M. Wearable inertial wensors for human motion analysis: a review. *IEEE Sensors Journal* **PP** (2016).
16. Luo, Y. *et al.* A database of human gait performance on irregular and uneven surfaces collected by wearable sensors. *Scientific data* **7**, 1–9 (2020).
17. Saudabayev, A., Rysbek, Z., Khassenova, R. & Varol, H. A. Human grasping database for activities of daily living with depth, color and kinematic data streams. *Scientific data* **5**, 1–13 (2018).
18. García-de-Villa, S., Casillas-Pérez, D., Jiménez-Martín, A. & García-Domínguez, J. J. Simultaneous exercise recognition and evaluation in prescribed routines: Approach to virtual coaches. *Expert Systems with Applications* **199**, 116990 (2022).
19. Lin, J. F. & Kulić, D. Human pose recovery using wireless inertial measurement units. *Physiological Measurement* **33**, 2099–2115 (2012).
20. Morrow, M. M. B. *et al.* Validation of inertial measurement units for upper body kinematics. *Journal of Applied Biomechanics* (2016).
21. Allseits, E. *et al.* The development and concurrent validity of a real-time algorithm for temporal gait analysis using inertial measurement units. *Journal of Biomechanics* **55**, 27–33 (2017).
22. Müller, P., Bégin, M.-A. S. T. & Seel, T. Alignment-free, self-calibrating elbow angles measurement using inertial sensors. *IEEE Journal of Biomedical and Health Informatics* **21**, 312–319 (2017).
23. X-io technologies. NGIMU. https://x-io.co.uk/ngimu/ (2020).
24. Optitrack. Motive: optical motion capture software. https://optitrack.com/software/motive/ (2020).
25. Casas-Herrero, A. Effect of a multicomponent exercise programme (vivifrail) on functional capacity in frail community elders with cognitive decline: study protocol for a randomized multicentre control trial. *Trials* **20**, 362 (2019).
26. Casas-Herrero, A. *et al.* Effects of vivifrail multicomponent intervention on functional capacity: a multicentre, randomized controlled trial. *Journal of Cachexia, Sarcopenia and Muscle* (2022).
27. Romero-García, M., López-Rodríguez, G., Henao-Morán, S., González-Unzaga, M. & Galván, M. Effect of a Multicomponent Exercise Program (VIVIFRAIL) on Functional Capacity in Elderly Ambulatory: A Non-Randomized Clinical Trial in Mexican Women with Dynapenia. *Journal of Nutrition, Health and Aging* **25**, 148–154 (2021).
28. Xu, C., He, J., Zhang, X., Yao, C. & Tseng, P.-H. Geometrical kinematic modeling on human motion using method of multi-sensor fusion. *Information Fusion* **41**, 243–254 (2018).
29. Crabolu, M. *et al. In vivo* estimation of the shoulder joint center of rotation using magneto-inertial sensors: MRI-based accuracy and repeatability assessment. *BioMedical Engineering Online* **16**, 1–18 (2017).
30. García-de-Villa, S., Jiménez-Martín, A. & García-Domínguez, J. J. Novel IMU-based adaptive estimator of the center of rotation of joints for movement analysis. *IEEE Transactions on Instrumentation and Measurement* **70**, 1–11 (2021).
31. Frick, E. & Rahmatalla, S. Joint center estimation using single-frame optimization: Part 1: numerical simulation. *Sensors (Switzerland)* **18**, 1–17 (2018).
32. Seel, T. & Schauer, T. Joint axis and position estimation from inertial measurement data by exploiting kinematic constraints. *2012 IEEE International Conference on Control Applications (CCA)* 0–4 (2012).
33. Crabolu, M., Pani, D., Raffo, L., Conti, M. & Cereatti, A. Functional estimation of bony segment lengths using magneto-inertial sensing: Application to the humerus. *PLoS ONE* **13**, 1–11 (2018).
34. Mathworks. MATLAB & Simulink. https://in.mathworks.com/ (2020).
35. Optitrack. NaturalPoint product documentation, see 2.2. https://v22.wiki.optitrack.com/ (2020).
36. García-de-Villa, S., Jiménez-Martín, A. & García-Domínguez, J. J. A database of physical therapy exercises with variability of execution collected by wearable sensors. *Zenodo* https://doi.org/10.5281/zenodo.6319979 (2022).
37. Leardini, A. *et al.* A new anatomically based protocol for gait analysis in children. *Gait and Posture* **26**, 560–571 (2007).
38. Lencioni, T., Carpinella, I., Rabuffetti, M., Marzegan, A. & Ferrarin, M. Human kinematic, kinetic and EMG data during different walking and stair ascending and descending tasks. *Scientific Data* **6** (2019).

39. Kwolek, B. *et al.* Calibrated and synchronized multi-view video and motion capture dataset for evaluation of gait recognition. *Multimedia Tools and Applications* **78**, 32437–32465 (2019).
40. Açici, K., Erdaş, Ç. B., Aşuroğlu, T. & Oğul, H. HANDY: A benchmark dataset for context-awareness via wrist-worn motion sensors. *Data* **3**, 1–11 (2018).
41. Roda-Sales, A., Vergara, M., Sancho-Bru, J. L., Gracia-Ibáñez, V. & Jarque-Bou, N. J. Human hand kinematic data during feeding and cooking tasks. *Scientific Data* **6**, 1–10 (2019).
42. Jarque-Bou, N. J., Vergara, M., Sancho-Bru, J. L., Gracia-Ibáñez, V. & Roda-Sales, A. A calibrated database of kinematics and EMG of the forearm and hand during activities of daily living. *Scientific Data* **6**, 1–11 (2019).
43. Reiss, A. & Stricker, D. Creating and benchmarking a new dataset for physical activity monitoring. *ACM International Conference Proceeding Series* (2012).
44. Finocchietti, S., Gori, M. & Souza Oliveira, A. Kinematic profile of visually impaired football players during specific sports actions. *Scientific Reports* **9**, 1–8 (2019).
45. Szczęsna, A., Błaszczyszyn, M. & Pawlyta, M. Optical motion capture dataset of selected techniques in beginner and advanced Kyokushin karate athletes. *Scientific Data* **8**, 2–8 (2021).
46. Vakanski, A., Jun, H. P., Paul, D. & Baker, R. A data set of human body movements for physical rehabilitation exercises. *Data* **3** (2018).
47. Ar, I. & Akgul, Y. S. A computerized recognition system for the home-based physiotherapy exercises using an RGBD camera. *IEEE Transactions on Neural Systems and Rehabilitation Engineering* **22**, 1160–1171 (2014).
48. Preatoni, E., Nodari, S. & Lopomo, N. F. Supervised machine learning applied to wearable sensor data can accurately classify functional fitness exercises within a continuous workout. *Frontiers in Bioengineering and Biotechnology* **8** (2020).
49. Opensim. https://opensim.stanford.edu/ (2020).
50. Kianifar, R., Lee, A., Raina, S. & Kulic, D. Automated assessment of dynamic knee valgus and risk of knee injury during the single leg squat. *IEEE Journal of Translational Engineering in Health and Medicine* **5** (2017).
51. Vecteezy. Free Vector Art. https://www.vecteezy.com/ (2022).

## Acknowledgements

The authors would like to thank Andrea MartÃnez-Parra and F. Javier Redondo-GarcÃa for their collaboration in the measurement campaign and data processing, and the volunteers that performed the exercises to create this database. This work was supported by Junta de Comunidades de Castilla La Mancha (FrailCheck SBPLY/17/180501/000392), the Spanish Ministry of Science, Innovation and Universities (MICROCEBUS RTI2018-095168-B-C51) and Comunidad de Madrid (RACC CM/JIN/2021-016).

## Author contributions

S.G.V., A.J.M. and J.J.G.D. conceived the experiments, S.G.V. conducted the experiments and processed the data, and together with A.J.M. and J.J.G.D. analyzed the results. All authors reviewed the manuscript.

## Competing interests

The authors declare no competing interests.

## Additional information

**Correspondence** and requests for materials should be addressed to S.G.-d-V.

**Reprints and permissions information** is available at www.nature.com/reprints.

**Publisher's note** Springer Nature remains neutral with regard to jurisdictional claims in published maps and institutional affiliations.